\documentclass[runningheads]{llncs}
\usepackage[T1]{fontenc}
\usepackage{graphicx}
\usepackage{booktabs}
\usepackage{courier}
\usepackage{comment}
\usepackage{diagbox}
\usepackage{colortbl} 
\usepackage{float}
\usepackage{xcolor} 
\begin{document}
%
% For ANTS 2024, full-length papers are strictly limited to 11 pages + references, 
% using this template. This page limit include figures, tables, and all 
% supplementary sections (e.g., Acknowledgements). The only exclusion from 
% these page limits is the reference list, which should have an appropriate 
% length with respect to the state of the art.
%
\title{A Risk Estimation Study of Native Code
Vulnerabilities in Android Applications}
% Title must be capitalized according to standard 'Title Case' style (i.e., all 
% words should be capitalized, except for articles, prepositions, and conjunctions).
%
% Any acknowledgements should be located at the end of the paper, in a final 
% Acknowledgements subsubsection.
% Do not format acknowledgments as a footnote, anywhere in the paper.
%
\titlerunning{Risk Algorithm Native Code Vulnerabilities Android}
% an abbreviated paper title here.
% The abbreviated title must also be capitalized according to standard 'Title Case'
% style (i.e., all words should be capitalized, except for articles, prepositions,
% and conjunctions).
% Do not use a \newline command in the abbreviated title.
%
%
\author{Silvia Lucia Sanna\inst{1}\orcidID{0009-0002-8269-9777} \and
Diego Soi\inst{1}\orcidID{0009-0009-0092-9067} \and
Davide Maiorca\inst{1}\orcidID{0000-0003-2640-4663} \and
Giorgio Fumera\inst{1}\orcidID{0000-0001-5300-226X} \and
Giorgio Giacinto\inst{1}\orcidID{0000-0002-5759-3017}}
% Follow the naming convention in which the surname is the last name.
% In this field, provide the full first names (not only the initials).
% Do not include academic titles (e.g., Prof. or Dr.).
% Springer encourages the inclusion of author ORCIDs.
%
\authorrunning{F. Author et al.}
% In this field, give the initial of the first name(s) and the full surname.
% Give the first author's name. If there are precisely two authors, then
% give both the first and second authors' names (sample: {F. Author and S. Author}).
% If there are more than two authors, use 'et al.' after the name of the first author.
%
\institute{Department of Electrical and Electronic Engineering, University of Cagliari, Italy}
% Author affiliation information should include the following, using the 
% \institute{} and \email{} fields: department, faculty, university, 
% company (if applicable), city, country, and email address. Do not include 
% the street address or ZIP code (ANTS 2024 does not use a postal address).
% The email address of the corresponding author is mandatory to include.
%
\index{SurnameAuthor1, FirstnameAuthor1}
\index{SurnameAuthor2, FirstnameAuthor2}
\index{SurnameAuthor3, FirstnameAuthor3}
% After the \institute{} entries, include an \index{} entry for each author, 
% giving the full surname, followed by the full first name(s).
%
\maketitle              % typeset the header of the contribution
\begin{abstract}
Android is the most used Operating System worldwide for mobile devices, with hundreds of thousands of apps downloaded daily. Although these apps are primarily written in Java and Kotlin, advanced functionalities such as graphics or cryptography are provided through native C/C++ libraries. These libraries can be affected by common vulnerabilities in C/C++ code (e.g., memory errors such as buffer overflow), through which attackers can read/modify data or execute arbitrary code. The detection and assessment of vulnerabilities in Android native code have only been recently explored by previous research work. In this paper, we propose a fast risk-based approach that provides a risk score related to the native part of an Android application. In this way, before an app is released, the developer can check if the app may contain vulnerabilities in the Native Code and, if present, patch them to publish a more secure application. To this end, we first use fast regular expressions to detect library versions and possible vulnerable functions. Then, we apply scores extracted from a vulnerability database to the analyzed application, thus obtaining a risk score representative of the whole app. We demonstrate the validity of our approach by performing a large-scale analysis on more than $100,000$ applications (but only $40\%$ contained native code) and $15$ popular libraries carrying known vulnerabilities. The attained results show that many applications contain well-known vulnerabilities that miscreants can potentially exploit, posing serious concerns about the security of the whole Android applications landscape.
\end{abstract}

\section{Introduction}
\label{sec:intro}
The usage of mobile devices is increasingly growing due to their continuous advancements that allow people to carry out very different tasks, from surfing the internet to accessing banking or medical accounts. Smartphones are also extensively employed as multimedia devices (e.g., to watch movies or play games) and as aids for payment authentication and Public Administration services. Unfortunately, this variety of usage allows attackers to exploit vulnerabilities (by resorting to, e.g., phishing emails and messages or by exploiting memory errors) to take control of the target devices.

Among the various Operating Systems available for mobile devices, Android is the most used worldwide~\cite{mobileos}, and many of its applications can feature hundreds of millions of downloads. These apps often need to interact with native activities and components (e.g., camera and microphone) available through Native Code (typically C/C++) implementation, which may be written from scratch or taken from third-party libraries such as \texttt{Libpng} and \texttt{OpenCV}. For brevity, we refer to native third-party libraries as \emph{products}. In most cases, developers use publicly available libraries such as \texttt{Libpng} (for image management) by importing them into their projects.  As native libraries are written in memory-unsafe languages, they can suffer from typical vulnerabilities caused by wrong source code programming or design. Improperly managing pointers, arrays, and API calls can lead to overflow attacks or other vulnerabilities. A simple example of possible memory errors is \emph{buffer overflow}, which allows an attacker to send an input whose size is larger than required, thus writing data outside bounds and causing unpredictable behaviours. Exploiting vulnerabilities in native libraries can affect the functionality of the whole application, leading to some data exposure or, in the worst cases, to the loss of control of the device. For this reason, it is essential to manage the security of the used libraries when developing an Android application. Previous research works have only recently pointed out the need for better native code safety and vulnerability analysis~\cite{Almanee21_ICSE}.

However, finding and analyzing vulnerabilities is a very time and resource-consuming task requiring in-depth static and dynamic analysis of the native layer and its interaction with the Java/Kotlin code~\cite{Almanee21_ICSE}. Recent works also showed several validation problems related to the effective \emph{reachability} of vulnerable functions~\cite{Borzachiello22_ESORICS}. These issues may discourage analyzing the native layer security in their apps, thus often overlooking even well-known issues of public products. We propose a probabilistic approach that vulnerability researchers can use to have a first basic idea of the vulnerability to be checked manually. Our vulnerability detection on Android Native Code can be included in the process of producing and maintaining the Software Bill Of Materials (SBOM),  a detailed inventory of software components and their ingredients essential in software security and supply chain risk management (as described by the American Cybersecurity and Infrastructure Security Agency~\cite{cisasbom}). Different organizations worked on that, such as NIST~\cite{sbomnist}, who released in February $2022$ guidelines to be followed by developers and companies as a means of cyberattack prevention. In fact, SBOM has been introduced to provide guidance on the level of risk associated with the software, whether stand-alone or integrated into systems (such as in the case of Android Native Code). SBOM defines the most dangerous vulnerabilities and gives a global risk indication of the software vulnerabilities, stating the components with a greater likelihood of being affected (as in our methodology). 

This paper proposes an alternative strategy for Native Code vulnerability identification that does not involve resource-heavy analyses but leverages on \textit{public knowledge of known issues}. The idea is to yield a quick, lightweight approach that gives an idea of an application's possible \textit{known risks} to take immediate actions to address them. This is done through a \textit{a risk assessment algorithm} that leverages a combination of quick code analysis and public domain knowledge to provide a score of possible dangerousness of the application based on the vulnerabilities found.  

More specifically, our contributions can be summarised as follows: \emph{(i)} we propose a \emph{minimal complexity} Native Code analysis strategy oriented to the search for known vulnerabilities and issues by leveraging \emph{public domain knowledge}; \emph{(ii)} we define a \emph{risk assessment algorithm} that provides a dangerousness score that can aid security researchers to take immediate actions to patch the analysed applications; \emph{(iii)} we evaluate our methodology through a large scale analysis on $100,000$ APKs taken from the widespread application repository Androzoo\footnote{https://androzoo.uni.lu/}, but results are focused on $38,348$ apks which are those using at least one native library. To the best of our knowledge, no risk assessment algorithm or methodology has been published for vulnerabilities in the Native Code. The results attained in this paper demonstrate that a risk-based approach can be strongly beneficial in swiftly assessing vulnerabilities in Android applications, thus addressing this problem by working on their early detection and prevention.

The remainder of this paper is structured as follows. \textit{Section~\ref{sec:background}} presents a technical background about Android applications structure and vulnerabilities. Previous research is illustrated in \textit{Section~\ref{sec:soa}}, while the applied methodology is presented in \textit{Section~\ref{sec:methodology}}. Results are reported in \textit{Section~\ref{sec:results}}.  Finally, \textit{Section ~\ref{sec:conclusions}} discusses the limits and the future works that may be conducted to improve this work.

\section{Technical Background}
\label{sec:background}

Before introducing our methodology, some concepts need a brief explanation to provide the reader with basic knowledge about the core elements of Android applications and Native Code.
\subsection{ARM}
\label{sec:background:subsec:arm}
ARM, the acronym for \textbf{Advanced RISC  Machine}\footnote{https://www.arm.com/}, is the hardware architecture on which Android OS and apps are executed. It is commonly implemented in embedded systems, where developers design and sell the processor's architecture to vendors such as Samsung, Lenovo, and Oppo. 
ARM is based on \textbf{RISC} (i.e. Reduced Instruction Set Computer), an architecture with a smaller instruction set than x$86$/$64$ but with more general purpose registers and a load/store mechanism.
As an example, to modify a value of a register, it is required to move the value to the register (with the instruction \texttt{load}), make the desired arithmetic operations, and save it back to memory (with the instruction \texttt{store}). 

Regarding arithmetic operations, the architecture reduces branch complexity and number of instructions by supporting conditional execution (\texttt{less than equal}, \texttt{greater than equal}) and barrel instructions (\texttt{shift} and \texttt{rotation}). ARM is also useful for implementing co-processors by allowing the execution of different tasks to different cores of one processor. Hence, the program execution time is inversely proportional to the number of cores.

\subsection{Android OS}
\label{sec:background:subsec:androidos}
The proprietary open-source Android OS~\cite{androidos}, published by Google in the early 2000s, is the operating system running on ARM hardware and on which apps are built. Android mostly features six main layers~\cite{androidarch}: \emph{(i)} \textbf{Android System Apps}, featuring apps for standard activities (e.g. SMS, calendar, emails); \emph{(ii)} \textbf{Java API framework} containing Android APIs to make different software components communicate with each other; 
\emph{(iii)} \textbf{Native Libraries and core system services} written in C/C++ to manage activities and interact with physical device components; \emph{(iv)} \textbf{Android Runtime} to manage runtime for executing Android apps since Android $5.0$;
\emph{(v)} \textbf{Hardware Abstraction Layer} (HAL), which is a software-hardware interaction layer that employs specific hardware interface description language (HIDL), allowing detachment between OS and drivers (autonomous upgrade); \emph{(vi)} \textbf{Linux kernel}, based on an upgraded version of Linux kernel to such platform.

An interesting characteristic of Android OS is the permission level. In \emph{low-level} mode, users and groups can access file systems and specific resources. Conversely, permissions are restricted in \emph{high-level} mode, and apps are installed.

Android applications and framework layers are executed in Android RunTime (ART). ART is the runtime system that executes Dalvik Executable format and Dalvik bytecode. Since Android 5.0 Lollipop, it replaces the \textbf{Dalvik Virtual Machine} (DVM), a register and Java-based virtual machine designed to give an efficient abstraction layer to the OS. With the DVM, developers had to partially compile the app, while the DVM did the other parts at runtime. Instead, in Android RunTime (ART), Dalvik bytecode is compiled for ARM assembly during installation, and the app can run directly by executing machine instructions.

\subsection{Android apps}
\label{sec:background:subsec:androidapps}
As described in the previous paragraph, Android apps~\cite{androidapps} work in the Android application layer, which allows developers to extend OS functionalities without altering lower levels. To build an Android app, developers write source code in Java/Kotlin that is compiled to \texttt{.class} files. Then, these are translated to Dalvik bytecode and compiled to a single \texttt{.dex} file (Dalvik Executable), which is optimized, loaded, and interpreted by the Dalvik Virtual Machine when the app is run. 
Once the app is compiled, it is packed as an \texttt{.APK} file, a sort of zip archive containing all files needed for its execution, and structured as follows:
\begin{itemize}
    \item \texttt{\textbf{AndroidManifest.xml}}, a file containing components and permissions needed by the app;
    \item \texttt{\textbf{classes.dex}} containing the assembled source code (i.e. list of classes, methods, and bytecode); %a disassembler is required for its analysis.
    \item \texttt{\textbf{res}}, a directory containing all elements related to the visual presentation;
    \item \texttt{\textbf{assets}}, a directory containing external resources used by the app while under execution;
    \item \texttt{\textbf{META-INF}}, a directory with the files used by the Java platform to interpret and configure the app;
    \item \texttt{\textbf{lib}}, a directory containing all platform-dependent native libraries written in C/C++ and compiled in the different \textbf{ABIs} (Application Binary Interface\footnote{An application binary interface (ABI) is an interface between the operating system and its applications. Each ABI is defined in Android by the combination of CPU and instruction set because each device uses its own.}) found in different sub-directories.
\end{itemize}
Each app is also made of four different active asynchronous component types. These are the \textit{activities}, entry points for user interaction;  \textit{services}, general-purpose entry points to keep the app running in the background; \textit{broadcast receivers}, intercommunication activities between the apps and the system; \textit{content providers}, whose aim is to allow apps to store and share data in the file system.

\subsection{Native Libraries}
\label{sec:background:subsec:nativelib}

As explained before, developers use Native Code to interact with native components and hardware. Moreover, developers can use Android NDK (Native Developer Kit~\cite{androidndk}) to import third-party native libraries without re-implementing them. 
These libraries are typically compiled C/C++ code with two kinds of extensions: \texttt{.so}, which can be defined as native shared libraries; \texttt{.a}, namely native static libraries, which are those linked to others.

Once compiled, the libraries are \textbf{ELF} (Executable and Linkable Format) files like the ones resulting from compiling a C code for Linux. The three most essential headers are \emph{(i)} \emph{ELF header}, with the \emph{magic number} to recognize the file format, compilation architecture, version, and information about sections; \emph{(ii)} \emph{program header} to create a process image; \emph{(iii)} \emph{section header} containing all the file's sections (\texttt{.bss}, \texttt{.data}, \texttt{.text}, etc.).

The analysis of an ELF can be performed with several command line tools (e.g. \emph{readelf} and \emph{elfdump}), reverse engineering software such as IDAPro and Ghidra, or python libraries like pwntools, which was employed in this work as explained later in Section~\ref{sec:methodology}. 

\subsection{Common Vulnerability Exposure}
\label{sec:background:subsec:cve}

Once a vulnerability is found in any application or hardware, it is often disclosed in a public dataset following the \textbf{CVE standard}~\cite{cve} with a name featuring the form \emph{CVE-DisclosureYear-ProgressiveNumberForThatYear}. Each database entry contains the product (i.e. an extensive library known commercially, such as \texttt{Libpng} or \texttt{OpenCV}), the affected product vendors, the publication date, and a human-readable description. The latter is very important, as it contains information about the vulnerability, such as the name of the vulnerable product (with the version) and the name of the function that allows the attack to be carried out. Sometimes, the attack and procedure to patch the vulnerability are also described. Each CVE is also quantified with a score according to \textbf{CVSS} (Common Vulnerability Score System)~\cite{cvss}. The latter defines three metric groups:
\begin{itemize}
    \item[-] \emph{Base Metric}, a constant severity value over time and across the user platform. It is composed of the \textbf{Exploitability metric}, which expresses the ease and technical levels required to exploit the vulnerability, and the \textbf{Impact metric} that quantifies the damage due to a successful exploit;
    \item[-] \emph{Temporal Metric}, a severity value that considers vulnerability's changes over time but is constant across the user platform;
    \item[-] \emph{Environmental Metric}, which reflects severity scores depending on the user's environment, possibly considering the presence of defence systems and security controls to mitigate the consequences of an attack. 
\end{itemize}

At the moment of writing this paper, different vulnerabilities have been published for Android: about $6305$ vulnerabilities in Android OS by Google, more than $100$ by Samsung, and only $2$ for Motorola; only $3$ vulnerabilities regarding Android hardware. Additionally, some vulnerabilities directly concern Android applications (the majority of them regarding Java management). Notably, vulnerabilities in Android Native Code cannot be addressed easily and directly as the Native Code is mostly embedded in other third-party libraries (see later sections of this work, particularly in Table~\ref{sec:methodology:tab:products}).

Up to now, there are different public databases containing CVEs, such as Mitre CVE~\cite{mitre}, CVE Details~\cite{cvedetails}, and National Vulnerability Database by NIST (NVD)~\cite{nvd}. Most public databases about vulnerabilities are derived from Mitre, but they add some details and more technical information.

\subsection{Risk Assessment}
\label{sec:background:subsec:risk}
In this work, we developed a risk assessment algorithm for vulnerabilities in the Native Code of Android applications. We now provide a brief introduction to the topic.

\textit{Risk assessment} includes a set of techniques and methods to determine the risk of an asset in a specific scenario. It is not only used in cybersecurity, but it is a general concept that can be applied to almost every field where undesired events could affect and damage the system. Given a specific scenario and a potential threat to an asset, we have to evaluate the \emph{likelihood} of the threat damaging the asset\footnote{In cybersecurity, the asset is the technical infrastructure we have to protect from cybercriminals.} in a quantitative (with numbers and specific metric measure), semi-quantitative (with numbers without a specific metric measure) or qualitative (with specific terms) way. The threat is often considered a deliberate attack, depending on the attacker's capability and the infrastructure protection mechanisms. The damage, of course, is the loss of infrastructure when the attack is successful. Technically, the risk assessment procedure involves tests such as penetration tests and simulation attacks where the company assesses the likelihood of being attacked. In general, after a risk assessment study, the company takes countermeasures to improve the protection of the assets. Different standards have been published about risk assessment, such as ISO 27005:2008 \cite{iso} (the one we used in this work) and NIST 800-30.

\section{Related Works}
\label{sec:soa}

Identifying Android Native Code vulnerabilities is one of the emerging research fields of cybersecurity, and as far as we know, very few works have been published on this topic. The prominent one is \textit{Librarian}~\cite{Almanee21_ICSE}, where the authors studied vulnerabilities in the top $200$ Android apps downloaded from Google PlayStore from September 2013 to May 2020. They studied a new algorithm called \emph{bin2sim}, capable of extracting $6$ features from each ELF in the app. Additionally, by applying binary similarities techniques, the tool correctly identifies different libraries and versions, implementing a whitelist approach for vulnerability detection. Despite the accuracy of this tool, their methodology is very heavy from the point of view of computational complexity. Moreover, they only say whether the APK and the native library are vulnerable or not without assigning a risk score. Since it is one of the most recent works on this topic, we decided to compare our methodology with their results, as we also based the product selection on their criteria and considered the $\pm$ two years time elapsed to patch a vulnerability. As described in Section~\ref{sec:results:subsec:librarian}, we had to find a way to compare our risk score on their dataset with their result because of the lack of score in their results. 

Although the early versions of the CVSS score have not been designed as a metric for risk estimation, over the years, the metric evolved to provide a reliable measure to evaluate the impact of vulnerabilities and exploits. One of the seminal works that pointed out the weaknesses of the previous versions of CVSS was the paper by Allodi and Massacci~\cite{Massacci14_ACM}, who in $2014$ criticised the usage of pure CVSS base score without considering the presence of exploits in the wild for a given vulnerability. The authors proposed a novel way to include known attacks by merging CVEs published in NVD and exploits released on exploit-db (repository of computer software exploits and exploitable vulnerabilities), eits (black markets exploits), and sym (vulnerabilities exploited in the wild). By employing this methodology, they could assess the limitations of the CVSS score first version and reduce the risk sensitivity according to the known exploits of $45\%$. The third version of CVSS (v3.1, the one used in this work for the experimental part) is more consistent. A brand new version has been released during this writing: CVSS v4.0 reinforces the concept of CVSS as not just a mere base score, as it considers threats, environments, attack requirements, user interactions and other metrics focused on a more real CVSS value, as suggested by Allodi and Massacci. 

Other works addressed vulnerability detection. For example, Alves et al.~\cite{Alves16_IEEE} studied the correlation between software metrics and software vulnerabilities. The authors claim that metrics exist to identify bad software, which is also harder to verify and maintain, with unnoticed or inadvertently introduced vulnerabilities. The authors compiled $5750$ vulnerabilities from Linux Kernel, Mozilla, Xen Hypervisor, httpd, and glibc. Analyzing $2875$ security patches, they distinguished vulnerable and safe functions. The results emphasize early vulnerability management and the need for developers to use multiple metrics for predicting code vulnerabilities. Even Madeiros et al.~\cite{Medeiros17_IEEE} addressed this topic with a study on software metrics useful to detect security vulnerabilities in software development. They analyzed various software metrics, such as complexity and coupling metrics, as well as other structural quality indicators, and identified patterns and correlations indicating the presence of security vulnerabilities. The authors established a correlation between specific project-level metrics and the number of vulnerabilities present in the software systems. They also found a specific group of discriminative metrics different across the software systems but present in all of them and valuable to distinguish between vulnerable and non-vulnerable code. The software metrics were identified using a genetics algorithm and a random forest classifier. Instead, in~\cite{Xiao20_USENIX}, a method called MVP (Matching Vulnerabilities and Patches) has been presented to detect vulnerabilities using patch-enhanced vulnerability signatures with low false positive and false negative rates. This methodology can distinguish between already patched vulnerabilities and generate accurate vulnerability and patch signatures to improve vulnerability detection accuracy. Du et al. developed \texttt{LEOPARD}~\cite{Du19_ACM}, a framework in a lightweight approach to help security experts detect potentially vulnerable functions in a code base without prior knowledge of the known vulnerabilities. Leopard combines complexity and vulnerability metrics to identify potentially vulnerable functions, providing a more comprehensive vulnerability assessment. The vulnerability is detected at all levels of complexity without missing the low-complex ones. For this purpose, the authors used a binning-and-ranking approach, where functions are grouped into bins based on complexity metrics and then ranked within each bin using vulnerability metrics. The framework covers a substantial portion of vulnerable functions identified while only a fraction is flagged as potentially vulnerable, outperforming machine learning and static analysis methods.

Another interesting problem regarding vulnerability detection is \textbf{reachability}, i.e., analyzing whether or not a vulnerable function is called in the app during its execution. Borzachiello et al. proposed \textsc{DroidReach}~\cite{Borzachiello22_ESORICS}, a tool to detect reachable APIs using heuristic and symbolic execution. They were able to represent all possible paths a function may take within the \emph{Inter-procedural Control-Flow Graph} (ICFG), whose aim is to encode all paths starting from an application entry point. Due to the complex methodology introduced in their work and the high computational complexity, we did not implement it in our work but considered this case in the risk methodology. Instead, we highlighted the imported library issue in our work as they did.

Recently, Ruggia et al.~\cite{Ruggia22_TechRxiv} developed a new methodology to reverse engineer Android apps, focusing on identifying suspicious patterns related to native components. They used suspicious tags to train a Machine Learning algorithm for binary classification. In particular, they developed a static tool that analyzes the code blocks responsible for suspicious behaviours in detail. This work demonstrates the use of native code in malicious Android applications so that the analysis is more complex and the maliciousness is better achieved.

One of the first studies on Android Native Code exploits was proposed in $2013$ by Fedler et al.~\cite{Fedler13_ACM}. They introduced different techniques to provide various levels of protection against all known local root exploits without affecting the user experience. Their mitigation reduced the exploitability of Android devices. In those years, very few Android applications used Native Code, and their approach was unsuccessful in exploiting and targeting flaws in the Dalvik Virtual Machine. Nowadays, more applications use Native Code, and Android architecture has changed. It is also worth noting that even popular tools for Android APKs vulnerability detection, such as MobSF~\cite{OpenSecurity15_MobSF} and Qark~\cite{LinkedIn15_Qark}, and SEBASTiAn~\cite{Pagano23_TechRxiv} do not look for vulnerabilities in the Native Code, even if are good vulnerabilities detection tools.

Fuzzing is another technique to detect vulnerabilities, and recently, it has also been used in Android Native Code. One of the most popular fuzzers is AFL++, which has been adapted in Frida mode \cite{afl++frida} to interact and fuzz Android applications: detect the vulnerability in the C/C++ code and also check the interaction with the Java/Kotlin code. Other tools have been released, such as Android-AFL \cite{android-afl} and Libfuzzer \cite{libfuzzer}. All these tools are resource-consuming and sometimes could be more efficient in detecting. Different works \cite{Liu20_USENIX} are focusing on fuzzing Android Native components, but, as far as we know at the moment of writing this paper, none of them focuses on fuzzing Android application Native Code.

Android CVE analysis has been studied by Brant et al. in~\cite{Brant22_CNS} with a focus on the Android security bulletin within the last six years from $2022$. According to them, to have more secure Android systems, security bulletin updates must be designed with specific tests and improve code coverage of patched files. 
Only $13$\% of security bulletin updates contain fixed test files for that particular update, and among these, only $42.8\%$ has full patch coverage. Even if the percentage is still low, this is an interesting result, meaning that the community is beginning to address Android security and vulnerability detection.

\textit{LibRadar}~\cite{Ziang16_ICSE} demonstrated that a whitelist approach is inefficient because package names can be modified in many ways. For this reason, they released a detection tool based on stable code features that are obfuscation-resilient, such as APIs, which are also obfuscation-resilient. We tried to use their approach, but at the time of this writing, the tool was no more accessible.
Another interesting approach is the one proposed by Li et al.~\cite{Li17_ICSE}, which identifies libraries according to reference and inheritance relations between Java classes, methods, and other app metadata. Notably, a Native Library cannot be identified only according to Java interaction and inheritance, but specific C code syntaxes must be considered.

Other works such as \textit{GoingNative}~\cite{Afonso16_NDSS}, \textit{NativeGuard}~\cite{Sun14_CSPW} and \textit{AppCage}~\cite{Zhou15_SICC} focused only on the isolation and secure sandboxing of Native Code in Android applications by running the app in a protected but unrealistic environment. These methodologies are fundamental but cannot be considered in a real-world scenario where customized sandboxes are rarely employed. On the contrary, \textit{Ndroid}~\cite{Xue19_TIFS} and \textit{DroidNative}~\cite{Alam16_arXiv} focused on data flow between Java and Native Code and Native Code control flow patterns, also for malware detection. Finally, \textit{AdDetect}~\cite{Narayan14_ISSNIP} is a framework for advertisement library detection using semantic analysis with machine learning and hierarchical clustering techniques. The interaction between Native Code and Java Code is critical and must be considered, even if we are limited to C code vulnerability detection in our work.

As Section~\ref{sec:intro} mentions, most of these works feature high time and space computational complexity. For example, using Machine Learning approaches, an extensive dataset of samples is needed to train the model correctly, and the fine-tuning of the model can introduce additional complexities. On the other hand, the methodologies above in the literature have a good accuracy in the results. 
Instead, the methodology we propose in this paper needs very few resources; it is fast to execute even with a large-scale dataset and does not need any dataset on which to train the Machine Learning algorithms.

One of the scenarios for which this work has been developed is the SBOM, as explained in Section~\ref{sec:intro}. This concept has been well explained by Zahan et al.~\cite{Zahan23_IEEE}, where they discuss the importance of SBOM in improving cybersecurity, highlighting its benefits and challenges. The work has been based on the Log4Shell vulnerability: a zero-day remote code execution vulnerability discovered in Apache Log4j that significantly impacted software organizations in December $2021$ despite a few months earlier, EO (Executive Order) $14028$~\cite{NIST21_E014028} recognized SBOMs as a practice for enhancing software supply chain security, and NTIA released a report of minimum points to use SBOM in risk reduction. After the Log4Shell vulnerability, NIST recognized SBOM as one of the official practices organizations must follow to reduce cyber attacks and listed it in the first version of the Secure Software Development Framework. Moreover, the industries following SBOMs were able to identify the Log4Shell vulnerability rapidly and had a more effective response. 

In the work, Zahan et al. highlight the benefits of using SBOMs, which include risk management, vulnerability detection, supply chain transparency, proactive management of security risks, and effective response to cyber threats. As part of EO $14028$, SBOMs can improve the nation's cybersecurity, but it still has to be standardized in the industry, and some challenges still must be solved, such as the standardization of data requirements, the enhancement of solid guidelines and the practitioners' collaboration. 

\section{Methodology and Implementation Details}
\label{sec:methodology}
This section illustrates the methodology used for vulnerability detection in the Native Code of Android applications. First, we describe the creation of the purpose-specific database. Then, we detail the library extraction algorithm we developed to analyze the information from each APK studied. Lastly, the risk assessment algorithm is fully explained.

Given an APK, we extract the native library from its \textit{lib} directory. As the compiled library is an ELF file, we need specific reverse engineering tools and techniques to extract data for vulnerability detection. Such data is the list of functions and the product name, along with the version to which the library belongs. Once we have this information, we can match this result with a database of known vulnerabilities. The database is \emph{purpose-specific}, containing for each CVE a field with the affected vulnerable version and function. At the moment of this writing, there is no publicly available database for this purpose, and with this specific structure, a system can easily access and read it. 

We need a list of $N$ products to construct the database, which are those whose vulnerabilities we want to look for in the apps. The product matching process is a \emph{whitelist approach} that assumes that none of the Android app's developers has changed the library name (keeping the real one employed during compilation time in the NDK). Once we have a match between the library under analysis and the database, we can assign a risk assessment score to the CVE found in the analyzed library. Our purpose is to give a risk value to each library (according to the risk of each CVE found) and, consequently, to each app to provide an alarm to developers and security researchers. The following sections will explain in more detail the methodology.

\subsection{Building a custom CVE Database}
\label{sec:methodology:subsec:db}

A CVE database is essential to check if the extracted data from the analyzed library have been declared in a published vulnerability listing. We decided to employ a custom database to reduce the query time to a public online database. Hence, our local database is a dump of data contained in the online selected website of CVEs (e.g., NVD) and with fields organized according to the aim of the research. As explained in Section~\ref{sec:background:subsec:cve}, the essential information about a vulnerability is the product's name with the version and the vulnerable functions of the library. All this information may be found inside the human-readable description. It does not follow a specific syntax such as \textit{"In product P of version v.n there is a vulnerable function F."}, so we need to employ Natural Language Processing (NLP) techniques to process the description and extract the valuable data. Specifically, we employed Python nltk library\footnote{https://www.nltk.org/}, adapting it to our use case and syntax.

The \textbf{product name} is easily matched with one of the $N$ selected products, and if one is mentioned inside the description, we know that the CVE has been found in that specific product. This can be further confirmed by searching all vulnerabilities by product name in the public database. 

The \textbf{version of the library} affected by the CVE is always a number that may come after the word \textit{version}, the product name, or preceding words with the meaning of \textit{after} or \textit{before}. Some examples are described in Table~\ref{sec:methodology:tab:exampleDescription2version}: (i) if there is a word with the meaning of \textit{before} (i.e., before, prior, earlier, etc.), every product version lower than the one found in the description is vulnerable; (ii) if there is a word with the meaning of \textit{after} (i.e., after, following, successive) every product version higher than the one found in the description is vulnerable; (iii) if no word with the precedents meaning is found, the affected product version is only the one found in the description. 

\begin{table*}[]
    \centering
    \caption{CVE descriptions and corresponding vulnerable product version.}\label{sec:methodology:tab:exampleDescription2version}
    \begin{tabular}{@{}p{8cm}l@{}}
        \toprule
        \textbf{Description} & \textbf{Vulnerable product version}\\[5pt]\midrule
        \emph{In product P \textbf{before version v.n} the F function can be used for buffer overflow.} & every version $<=$ v.n is vulnerable \\[15pt]\midrule
        \emph{In product P \textbf{version v.n}, F function can be used for attack} &  version == v.n is vulnerable \\[5pt]\midrule
        \emph{In product P \textbf{version v.n and after}, F function can lead to buffer overflow.} & every version $>=$ v.n is vulnerable\\[15pt]
        \bottomrule
    \end{tabular}
    \label{tab:db}
\end{table*}

The last data to extract from the description is the \textbf{function name}. We rely our strategy on how programmers typically recognize function names and give names to functions. Then, by looking at some CVE descriptions, we noticed that the function is never declared as, for example, \textit{function F}, represented uniquely by its name. The function name usually contains specific symbols (i.e. \_, ::, (), etc.). Moreover, if the name is made up of more than one word, it is camel-cased. As an example, \texttt{makeSum} is made up of two words: make and Sum.
In other cases, the function is not found, which means that the whole product version is vulnerable, but no description of the vulnerable function is provided.

This methodology was developed in iterative steps with manual cross-validation to check if the algorithm worked correctly. 

\subsection{Library Extraction, Analysis and Association}
\label{sec:methodology:subsec:lib}

\begin{figure*}[]
    \centering
    \includegraphics[width=0.7\textwidth]{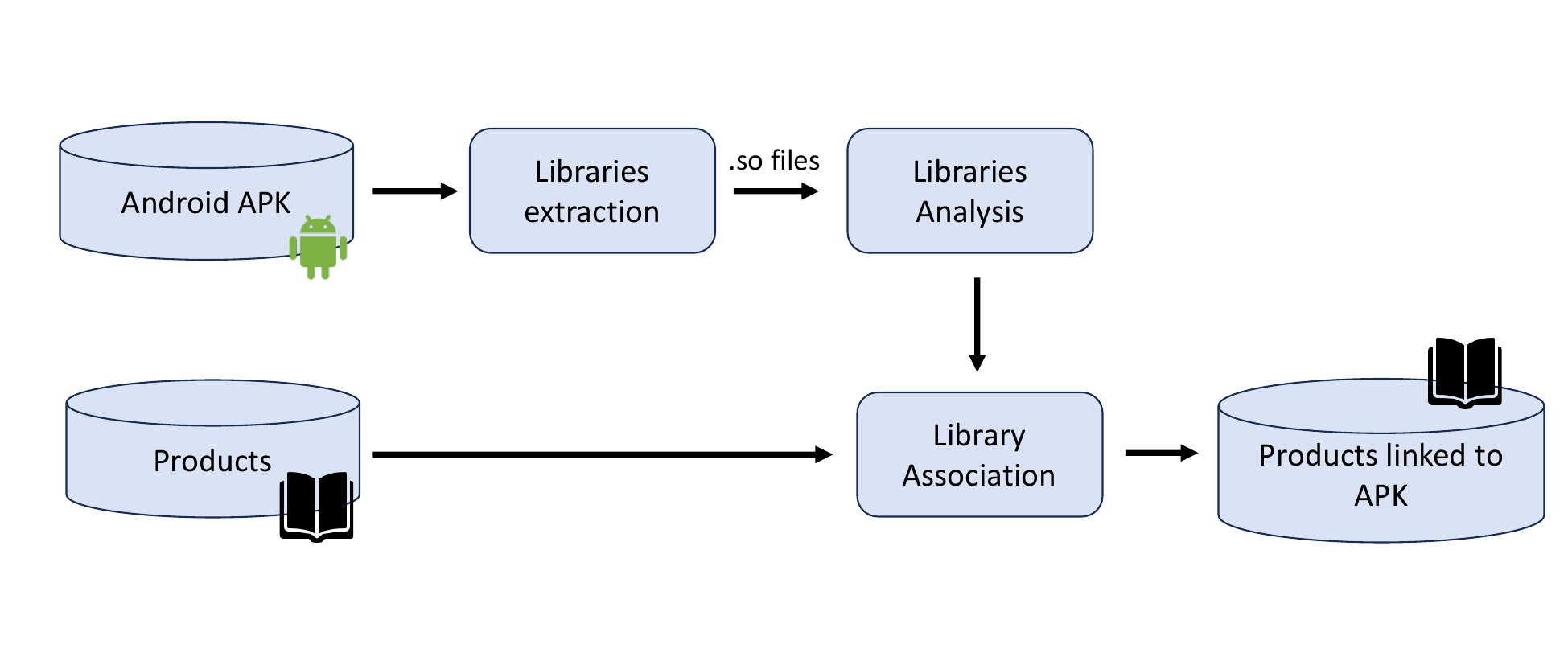}
    \caption{Workflow of the approach to extract, analyse and associate native libraries, specifically in the use-case of vulnerability researchers.}
    \label{fig:schema}
\end{figure*}

The overall approach, subdivided into library extraction, analysis, and association, is depicted in Figure~\ref{fig:schema}. The presented methodology can be used as it is by vulnerability researchers and adapted for developers to release a secure Android application.

\textbf{Library Extraction} is the first step of the analysis part, and, as the name states, it consists of the extraction of the compiled library (ELF file) from the Android application. It is done by unzipping \texttt{.so} files from the \texttt{lib} directory of each APK. Indeed, libraries are compiled according to different ABIs and saved inside the application. We can choose to extract libraries only for specific architectures or to analyze libraries for all available ABIs. In this work, we considered all ELF files from \texttt{armeabi-v7a} directory, and if not available, looked for \texttt{arm64-v8a} or \texttt{x86\_64} as they represent the most popular architectures.

\textbf{Library Analysis} is the part where we need to extract from the library the data for vulnerability detection. In particular, we need to know the list of functions in the library and the product name and the version to which the library belongs. This step can be done with different reverse-engineering tools, such as Ghidra and its Python extension for automation. Typically, the product name and its version can be retrieved in the strings section (.rodata section), while the defined functions in the ELF file can be retrieved from the Symbol Table section. In this work, we used pwntools~\cite{pwntools}, a popular Python framework for binary analysis and exploitation. Specifically, we employed its \texttt{ELF} module, which allows the analysis of ELF files from which the strings and the list of functions are extracted. The version is taken from the result of the \texttt{strings} Linux utility\footnote{\texttt{strings} tool in Linux is capable of retrieving all printable sequences of characters from the .rodata section of an ELF.}. At the same time, the functions are found inside the Symbol Table of the ELF without considering \texttt{.got} and \texttt{.plt} sections. A difficulty comes when binaries are stripped\footnote{A stripped binary is a binary without some debugging symbols and so with a lack of data.} where not all function names are available, or the names are unrecognizable.

\textbf{Library Association} is the part where each ELF file is associated with at least one of the selected $N$ products. This is a crucial task: if we do not link each ELF file to its related product, we cannot determine if the ELF file contains vulnerabilities, hence assigning a risk assessment score. Section~\ref{sec:soa} illustrates that different works used binary similarities techniques or Machine Learning algorithms; however, we decided to apply a simple identification algorithm because we know that every product uses a clear and unique syntax in strings and function names. For example, in \texttt{OpenCV}, we can find strings like \textit{General Configuration for OpenCV v.n} to declare the version and \textit{xxxx\_cv\_xxxx} in the function names. 
For this reason, if we find these syntaxes in strings and functions, we can link the analyzed ELF file to a product. 

Some binaries can be linked to multiple products due to imported libraries. Indeed, in a compiled library, there can be traces of the \emph{primary} library and the \emph{secondary} libraries (i.e., the ones employed by the primary). We considered the ELF files belonging to all the retrieved products in this case, as a proper distinction between primary and secondary libraries can be hard to carry out in practice (an issue that has also been highlighted by Borzachiello et al.~\cite{Borzachiello22_ESORICS}).

When developers use our methodology, thanks to the library association step, they can find vulnerabilities in the \textit{secondary} library (i.e. a library contained in the main library they are importing in the project as described here above) and take actions to mitigate them. Developers also use the library extraction and analysis part as we designed because when importing a third-party native library in an Android project for apk creation, Android Studio compiles the library, and to analyse it, developers have to extract and analyse the compiled ELF file as security analysts do. Instead, when developers want to check if the native library contains vulnerabilities before importing them into the Android Studio project to create the application, they can immediately check the database, find a version with fewer vulnerabilities, and patch them.

\subsection{Risk Assessment Algorithm}
\label{sec:methodology:subsec:risk}

We developed a risk assessment algorithm to give a risk value to each library (and consequently to each app) to provide an alarm to security researchers. Even though a CVE is present in the analyzed libraries within an ELF, we are not $100\%$ sure that it is exploitable due to stripped binaries, imported products, and without considering the reachability problem. Moreover, developers can rename the functions, patch their content or use obfuscation techniques. For this reason, we can approach the problem in \emph{probabilistic terms}, and we developed a semi-quantitative risk assessment algorithm.

According to \textbf{ISO 27005:2008}~\cite{iso}, the risk can be evaluated as the product between $3$ factors:
\begin{equation} \label{eq:1}
risk = threat * impact * vulnerability 
\end{equation}

A \textbf{threat} corresponds to an action that negatively impacts the device. Hence, the threat factor can be associated with the ease with which an attack can be carried out. To quantify it, we can use the \emph{CVSS exploitability} value since it has a similar definition, regardless of the attacker's capabilities. The \textbf{impact} is the damage caused to the system if the vulnerability is exploited. It can be quantified with the CVSS impact value without considering the architecture of the victim device. 

\begin{table*}[]
    \scriptsize
    \centering
    \caption{Risk matrix to determine the qualitative risk value.}
    \begin{tabular}{|p{2.5cm}|p{1.5cm}|p{1.5cm}| p{1.5cm} |p{1.5cm}| p{1.5cm}|}
    \hline \bf \diagbox[width=\dimexpr \textwidth/5+2\tabcolsep\relax, height=1.4cm]{Threat*Impact}{\rule{0pt}{2.3ex}Vulnerability}%Score (threat * impact)/ Vulnerability 
    & \bf None & \bf Low & \bf Medium & \bf High & \bf Critical\\ [10pt]
    
    \hline \rule{0pt}{3ex} \bf Critical & \cellcolor{yellow!50} Medium & \cellcolor{orange!50} High & \cellcolor{orange!50} High & \cellcolor{red!50} Critical &\cellcolor{red!50} Critical \\ [10pt]
    
    \hline \rule{0pt}{3ex} \bf High &\cellcolor{yellow!50} Medium & \cellcolor{yellow!50} Medium & \cellcolor{orange!50} High & \cellcolor{orange!50} High & \cellcolor{red!50} Critical \\ [10pt]
    
    \hline \rule{0pt}{3ex} \bf Medium & \cellcolor{green!50} Low & \cellcolor{yellow!50} Medium & \cellcolor{yellow!50} Medium & \cellcolor{orange!50} High & \cellcolor{orange!50} High \\ [10pt]
    
    \hline \rule{0pt}{3ex} \bf Low & \cellcolor{green!50} Low & \cellcolor{green!50} Low & \cellcolor{yellow!50} Medium & \cellcolor{yellow!50} Medium & \cellcolor{orange!50} High \\ [10pt]
    \hline
    \end{tabular} \\
    \label{tab:riskmatrix}
\end{table*}

\begin{itemize}
    \item \textbf{CRITICAL} when the CVE is present and surely exploitable;
    \item \textbf{HIGH} when a vulnerable library is found within the application, but we are uncertain about the CVE exploitability because the vulnerable API is not reachable, or we did not find the vulnerable function because the binary is stripped;
    \item \textbf{MEDIUM} when we can make the same assumptions of HIGH level for functions, but we cannot find the vulnerable version due to stripped binaries. Indeed, the library could be vulnerable because developers would be unaware of the dangers of function if the CVE were released after the app's publication. Moreover, apps falling in this category feature a difference between their release and the CVE publication dates, which are inferior to two years. According to Librarian~\cite{Almanee21_ICSE}, two years is the time developers use to apply a patch and mitigate vulnerability effectiveness after its release. Hence, within this period, it is very likely that the library would be affected.
    \item \textbf{LOW} when we cannot state whether the CVE is present. So, we establish a small level of risk when the found version and functions are not associated with any CVE;
    \item \textbf{NONE} when no native library is found, or the analyzed ELF files do not belong to our $N$ products.
\end{itemize}

Our study aims to establish a qualitative value of the risk. To do so, we have to rescale the semi-quantitative product between threat and impact (in a range between $0$ and $100$ as both of them have values between $0$ and $10$) into qualitative metrics, as detailed in Table~\ref{tab:quantitative}. The values have been determined according to the CVSS $3.1$ Qualitative Severity Rating Scale. 
Then, by applying the equation~\ref{eq:1}, we evaluate the risk according to the matrix in Table~\ref{tab:riskmatrix}. In this way, we have a qualitative risk assessment score to assign to each CVE present in each library of each Android application.

\begin{table}[h]
    \centering
    \caption{Mapping from semi-quantitative to qualitative values of the product between threat and impact. }\label{sec:methodology:tab:rescalingrisk}
    \begin{tabular}{@{}p{5cm}l@{}}
        \toprule
        \textbf{Semi-quantitative value} & \textbf{Qualitative value}\\[5pt]\midrule
        $90\div100$ & \emph{CRITICAL} \\[5pt]\midrule
        $89\div70$ & \emph{HIGH} \\[5pt]\midrule
        $69\div40$ & \emph{MEDIUM} \\[5pt]\midrule
        $39\div0$ & \emph{LOW} \\[5pt]
        \bottomrule
    \end{tabular}
    \label{tab:quantitative}
\end{table}

The purpose is to give a risk value to each library (and, consequently, to each app) and to provide swift alarms to security researchers. Once they are informed about the risks associated with the library or application, they can find a way to patch the vulnerability (e.g., upgrade the library to a non-vulnerable version, fix the library, etc.). To this aim, we assigned the highest score of the CVEs risk level to each library but also saved the affected CVEs in a log file. For example, if we find five CVEs with a LOW level and one with a MEDIUM level, we assign a MEDIUM risk to the library for a more effective alarm.
Another approach could be evaluating the average risk, giving the most representative risk: in the previous example, the result should be LOW, but in this way, we would underestimate the risk, which is far from our purpose. The same weighted approach has been adopted for the application risk level attribution.

\section{Results}
\label{sec:results}
In this Section, we illustrate the results of a large-scale analysis conducted on $100,000$ APKs from Androzoo. Additionally, to prove the efficiency of our methodology and risk algorithm and to make a comparison between our approach and Libriarian~\cite{Almanee21_ICSE}, we applied the approach to $32$ apps from the published dataset~\cite{LibrarianGitHub} by Almanee et al.

Since the Librarian dataset employs apps released between September $2013$ and May $2020$, we downloaded the updated versions of such apps (February $2023$) and whether the vulnerability risk changed in this amount of time.

\subsection{Dataset and products}
\label{sec:results:subsec:androzoo_products}

To apply the study to a large-scale dataset, we downloaded the first $100,000$ applications found in the Androzoo collection, which is a popular dataset developed by the University of Luxembourg with about $23$ million APKs dumped from different markets and years, also analyzed by various anti-malware engines. Among the $100,000$ applications, we selected only $38,348$, which contained native code (see Supplementary Data for the hash list).
In particular, since this study started in September $2021$, we downloaded the samples considering the list of apps from Androzoo csv.

Additionally, we built a dataset of products, i.e., a set of $N$ popular native libraries within the downloaded samples. This is a necessary step to associate each ELF file with at least one product, as expressed in Section~\ref{sec:methodology:subsec:lib}. Due to time and space constraints, we selected the most representative libraries within the $38,384$ APKs, considering the number of published vulnerabilities (CVE) for those libraries and the representativeness within the app chosen inside the large-scale dataset. 
So, in a similar fashion to what was made by Almanee et al.,~\cite{Almanee21_ICSE}, given the $38,384$ APKs we made a list of the native libraries' names, associated them to their own products, and computed statistics of the products considering the number of the published known vulnerabilities and their percentage of diffusion in the dataset (considering the top products to be selected). At the end, we had a dataset of $15$ products: \texttt{OpenCV, OpenSSL, FFmpeg, Libavcodec, Libavformat, Libswresample, Sqlite 3, LibWebp, Libpng, Libjpeg-turbo, Lua, Mono, Folly, Hermes, React-Native}. 

\begin{table*}[]
    \centering
    \caption{The Table shows the 15 selected products to perform the analysis, a brief description of their purpose, the number of released CVEs in the last five years, and the percentage of apps in our dataset that contain these products.}\label{sec:methodology:tab:products}
    \begin{tabular}{@{}lp{5cm}lll@{}}
        \toprule
        \textbf{Product} & \textbf{Description} & & \textbf{\# CVE} & \textbf{\% Dataset}\\[5pt]\midrule
        \texttt{OpenCV} & Real-time Computer Vision & & 34 & 5\\[5pt]\midrule
        \texttt{OpenSSL} &  Secure communication over network & & 232 & 1 \\[5pt]\midrule
        \texttt{FFmpeg} & Manage multimedia files (audio-video) & &  407 & 10 \\[5pt]\midrule
        \texttt{Libav\footnotemark[8]} & FFmpeg fork to manage multimedia files & & 106 & 5\\[5pt]\midrule
        \texttt{Sqlite 3} & Database engine & & 54 & 8 \\[5pt]\midrule
        \texttt{LibWebp} & Alternative to PNG, JPEG, GIF & & 14 & 5 \\[5pt]\midrule
        \texttt{Libpng} & Handle PNG images & & 44 & 5 \\[5pt]\midrule
        \texttt{Libjpeg-turbo} & Handle JPEG image format & & 1 & 16\\[5pt]\midrule
        \texttt{Lua} & Lua language interaction & & 15 & 3 \\[5pt]\midrule
        \texttt{Mono} & Create compatible tools with Framework .NET  & & 20 & 11\\[5pt]\midrule
        \texttt{Folly} & Core library components used by Facebook & & 3 & 11\\[5pt]\midrule
        \texttt{Hermes} & Fast startup of React Native apps & & 20 & 10\\[5pt]\midrule
        \texttt{React-Native} & Use React framework in applications & & 1 & 77\\[5pt]
        \bottomrule
    \end{tabular}
\label{tab:products}
\end{table*}

\subsection{Large-scale analysis Results}
\label{sec:results:subsec:androzoo}
Out of the $100,000$ downloaded APKs, about $40\%$ of them contained Native Code. In particular, there are $38,384$ APKs with at least one native library and a total of $225,638$ ELF files. Among these, $44,225$ belongs to at least one of our $15$ products. 
\begin{figure}[h]
    \centering
    \includegraphics[scale=0.4]{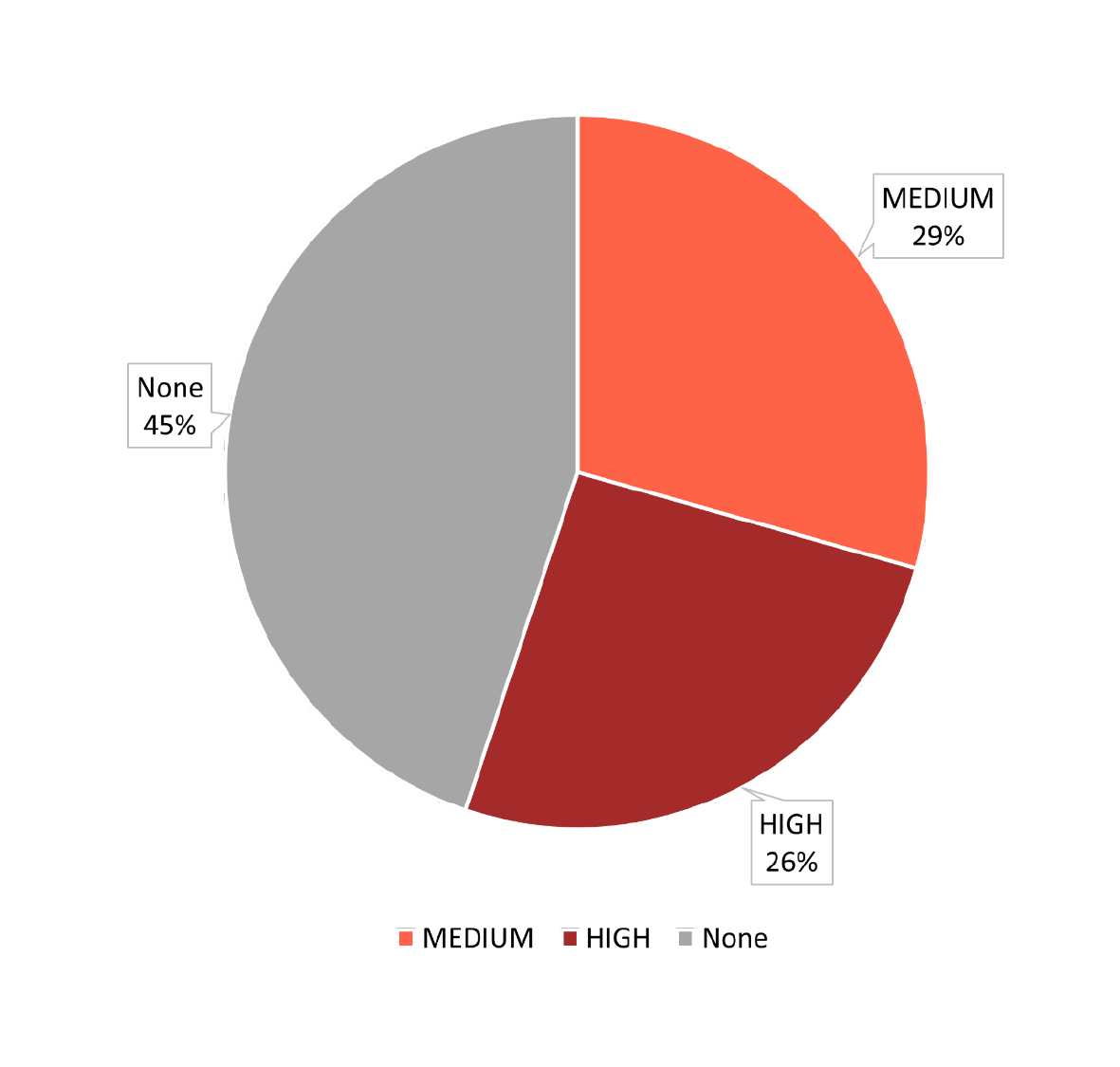 }
    \caption{The pie chart shows the percentage of apps for which a risk level has been computed by only identifying $15$ products. The NONE value means that the found Native Code does not belong to any of our $15$ selected products, but it can have vulnerabilities related to other libraries.}
    \label{sec:results:fig:androzoovuln}
\end{figure}
\begin{figure*}[]
    \centering
    \includegraphics[scale=0.20]{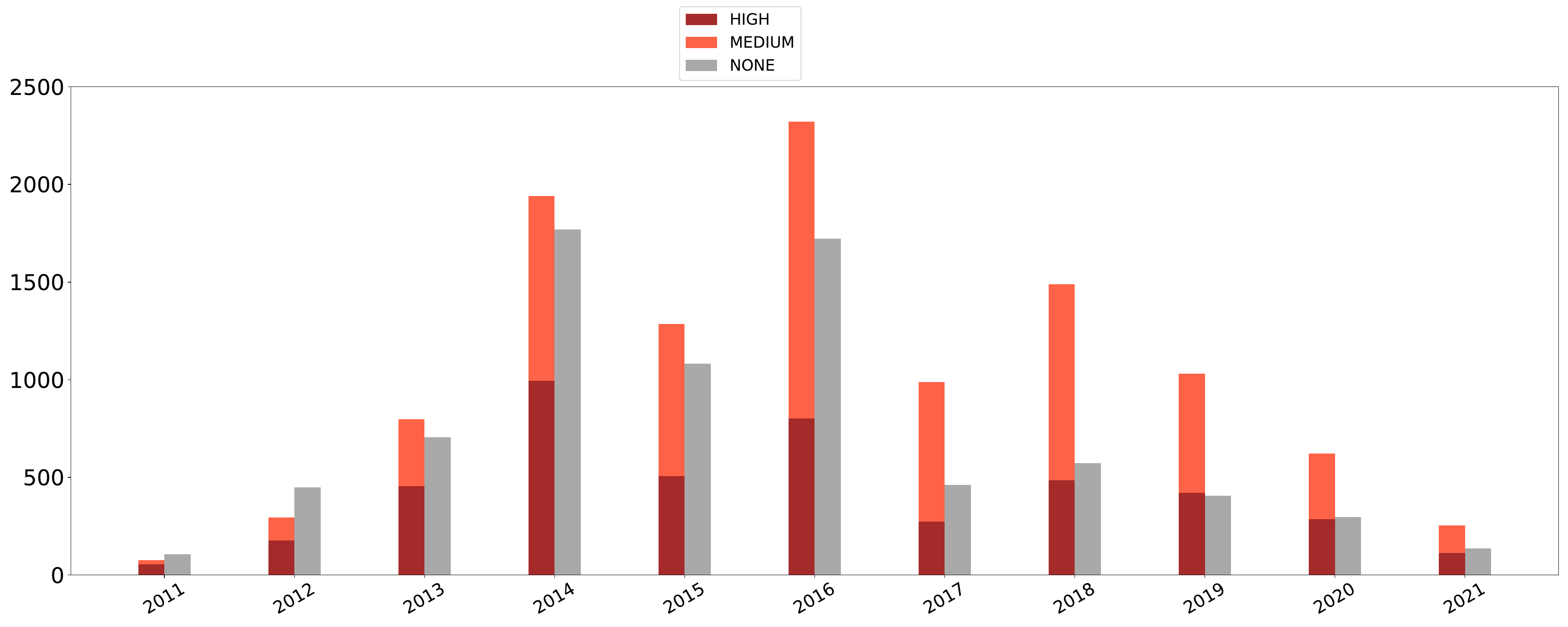}
    \caption{This histogram shows the risk level per year on the analyzed apps. Each year has 2 bars: red/left-bottom bar for HIGH risk, orange/left-upper bar for MEDIUM risk, and grey/right bar for NONE risk.}
    \label{sec:results:fig:androzooyears}
\end{figure*}
Regarding applications with Native Code, by considering only the products in the library-identification dataset as defined in Table~\ref{tab:products}, we could determine the risk for $55\%$ of them. Hence, about $24,000$ APKs belong to at least one of our $15$ products with a risk level \textit{HIGH-MEDIUM} as reported in Figure~\ref{sec:results:fig:androzoovuln}.

We also computed some statistics about the apps' year of release and belonging market. 

Figure~\ref{sec:results:fig:androzooyears} shows that the apps released from $2011$ to $2021$ have a medium risk. Note that the \textit{NONE} label does not mean that applications are not vulnerable at all, but that they belong to other products (i.e. no product among the selected ones is found). Expanding the analyzed products dataset may increase the number of vulnerable apps.

\begin{figure}[h]
   \centering
    \includegraphics[width=0.4\textwidth]{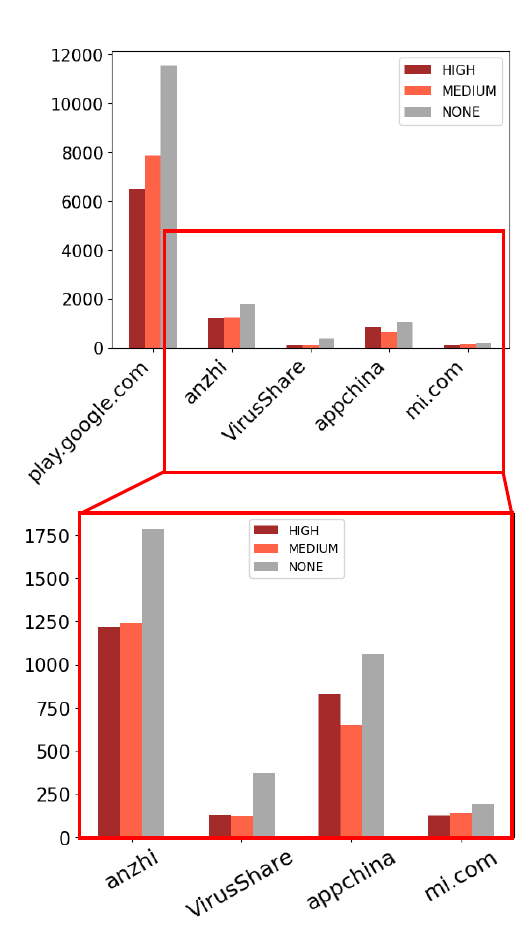}
    \caption{The histograms show the vulnerability risk levels (HIGH: red/left bar; MEDIUM: orange/central bar; NONE: grey/right bar) of the apps by markets. On the left, we plotted the main markets. On the right, we can see a detail of all markets except the Google Play Store.}\label{sec:results:fig:markets}
\end{figure}

\begin{figure*}[h]
    \centering
    \includegraphics[scale=0.15]{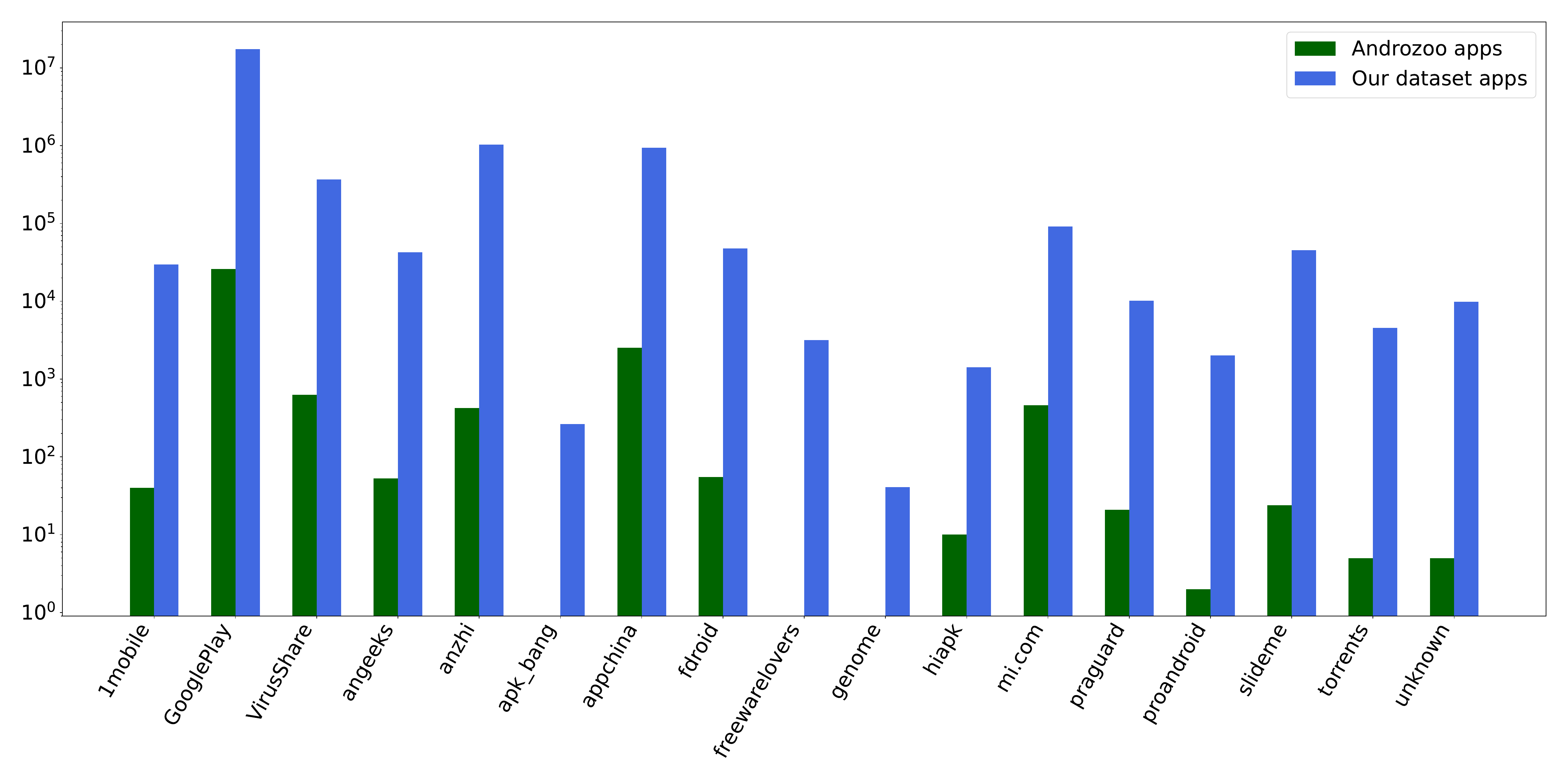}
    \caption{The histogram compares the number of apps for each market in the Androzoo dataset (green/left bar) and the apps used for vulnerability detection in our dataset (blue/right bar).}
    \label{sec:results:fig:marketsPerc}
\end{figure*}

The market with the most vulnerable applications is Google Play, as depicted in Figure~\ref{sec:results:fig:markets}. That is a reasonable result because most apps in the dataset are retrieved from the Google Play Store (Figure~\ref{sec:results:fig:marketsPerc}). The reason is that Google Play only limits its checks in understanding if an uploaded application can be classified as malware without considering vulnerabilities.

\subsection{Comparison with Librarian results}
\label{sec:results:subsec:librarian}
To demonstrate our methodology's effectiveness, we compared our results with the ones obtained by Almanee et al.~\cite{Almanee21_ICSE} by downloading their pubic dataset of 32 APKs~\cite{LibrarianGitHub} built in 2021. \footnotetext[8]{In our dataset \texttt{libav} consists of three libraries (i.e. \texttt{Libavcodec}, \texttt{Libavformat}, \texttt{Libswresample}) to code/decode, multiplex/demultiplex, and resample audio, and video.}We could infer almost the same library identification results by using the same products for library identification. 

While the methodology proposed by Almanee et al. was limited in saying whether the analyzed app was vulnerable (so only a result yes/no), our approach gives one of the presented risk levels in Section \ref{sec:methodology:subsec:risk}, we have to find a way to compare the results of the two methodologies. Hence, when our methodology gives a level to an application to which the other approach says is vulnerable, we gave the Librarian results in the same risk level set by our algorithm. On the other hand, when our methodology gives a level to an application, but the approach of Amanee et al. says it is not vulnerable, we gave the value of $0$ to the Librarian results. First of all, we infer the same results as the Librarian did for all applications, except for one which we detected as MEDIUM risk, but the Librarian says it has no vulnerabilities. From the results, we can affirm that $47\%$ of the apps have a HIGH risk level, $41\%$ MEDIUM risk level, and $12\%$ of the apps do not have vulnerable Native Code libraries (i.e., their products do not belong to our or Librarian dataset of products; the products do not contain CVEs).

% \begin{figure*}[]
%     \centering
%     \includegraphics[scale=0.25]{images/results/librarian.pdf}
%     \caption{The histogram shows the risk level for each of the Librarian apps given by our algorithm (left/red bar for each app) and the same value assigned by Librarian itself (right/orange bar).}
%     \label{sec:results:fig:librarian}
% \end{figure*}
\begin{figure*}[h]
    \centering
    \includegraphics[scale=0.15]{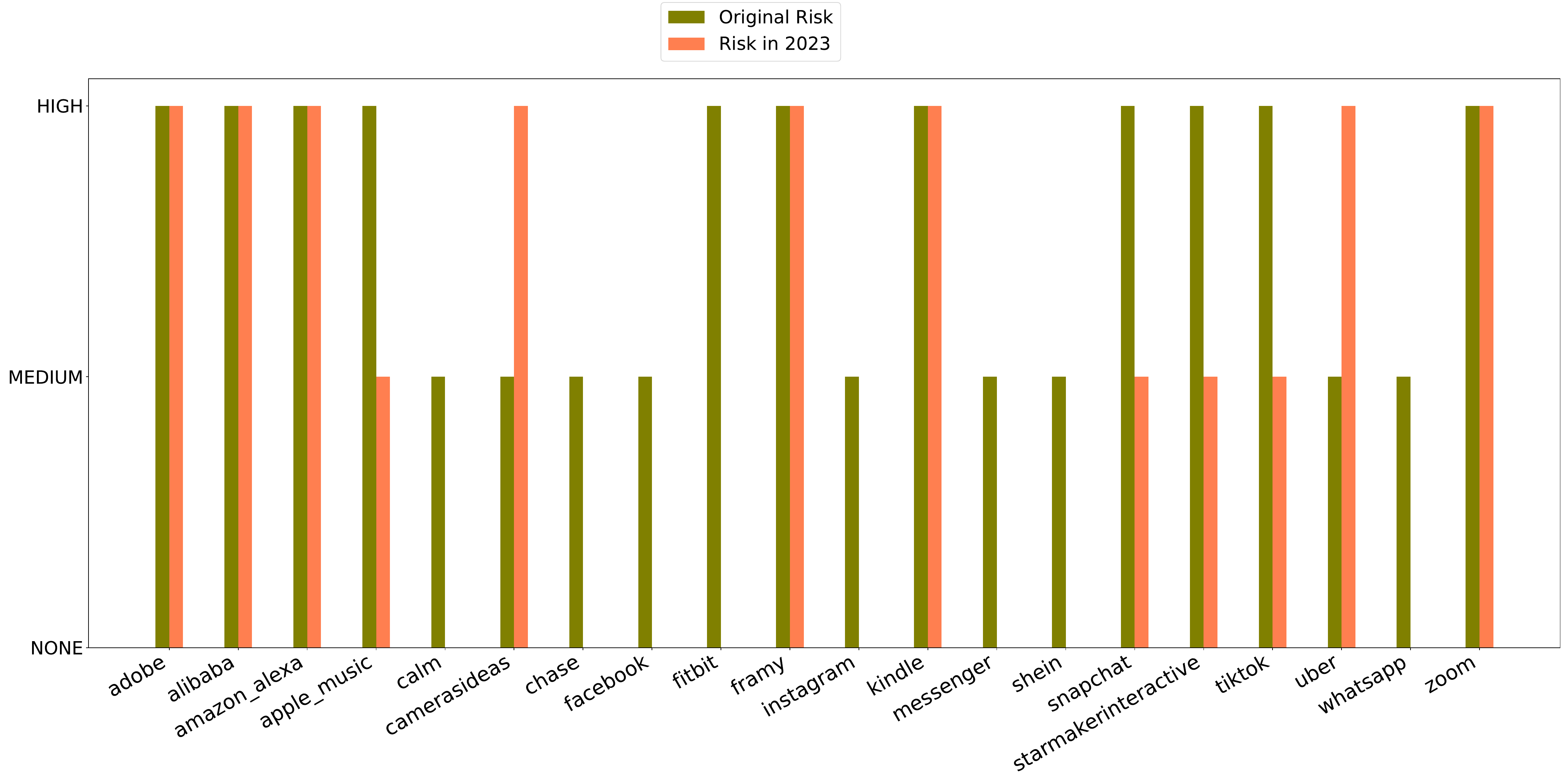}
    \caption{The histogram shows the risk level for the librarian apps (left/olive green bar) and the version on February $2023$ (right/coral bar).}
    \label{sec:results:fig:lastversion}
\end{figure*}

\subsection{Comparison with updated Librarian dataset}
\label{sec:results:subsec:lastversion}

The last experiment we performed was on the last version of the Librarian apps, downloaded on February 2023, used as a dataset to check whether the risk changed. 
As Figure~\ref{sec:results:fig:lastversion} shows, We can state that for the $55\%$ of the apps, the risk was reduced; for the $10\%$ of them, the risk remained constant, and for the $2\%$ of them, it increased. The null results are caused by Native Code that does not belong to any of the selected $15$ products. 

As seen in Figure~\ref{sec:results:fig:androzoovuln}, the selected 15 products are still insufficient to scan all apps, even though they are the most popular. Moreover, Facebook, Instagram, Messenger, and WhatsApp use Native Code developed by Meta instead of importing third parties as in the Librarian's version.

From the graph in Figure~\ref{sec:results:fig:androzooyears}, we have seen that the risk reduces over time, which is expected if we consider that various vulnerabilities have been addressed over the years. However, it is also interesting that \emph{the risk for various apps has not changed over time}, questioning the quality of vulnerability addressing in popular applications.

\section{Summary, Limitations and Future Works}
\label{sec:conclusions}
In this work, we proposed a simple and fast approach for vulnerability detection in Android Native Code by developing the first database of CVEs for vulnerability detection by easily accessing the vulnerable version of each app and its vulnerable functions. We combined this with developing a risk assessment algorithm for vulnerability management.

We demonstrated that even a simple approach like ours is efficient for vulnerable library identification, as we have been able to reproduce the same results as previous works by highlighting vulnerable applications on a much larger scale. Our methodology can aid developers and security researchers mitigate immediate risks by recommending fine-grained application patching, thus allowing them to release more secure Android applications in the different Android markets.

However, our methodology does not consider issues such as \textit{reachability} to determine if vulnerable functions are accessible in apps and \textit{stripped binaries} to assess the presence of the vulnerable function's name in the binary. In fact, our solution gives security researchers and developers a risk score so they can manually check the vulnerability. To better score the functions' reachability, security researchers can refer to other tools such as DroidReach~\cite{Borzachiello22_ESORICS}. Additionally, we do not check whether or not the vulnerable library has been patched. Indeed, even though the analyzed ELF file matches a vulnerable version or the function name, developers may have patched the function's body, for which binary similarities techniques must be used. Developers can also rename the functions or obfuscate their name (as well as the content). For these cases, our whitelist approach is insufficient to determine the risk and match if the found function is in the vulnerability database. All these issues can be addressed in future research works. 

In the future, we plan to address these issues by extending the product dataset and including as many libraries as possible to check how the risk changes. 
Concerning library identification, we plan to extract unique functions from each version of the products and use them as features for Machine Learning algorithms. 

\section{Competing interests}
The authors declare no competing interest.

% \section{Author contributions statement}

% Must include all authors, identified by initials, for example:
% S.R. and D.A. conceived the experiment(s),  S.R. conducted the experiment(s), S.R. and D.A. analysed the results.  S.R. and D.A. wrote and reviewed the manuscript.

\section{Acknowledgments}
This work was carried out while Silvia Lucia Sanna was enrolled in the Italian National Doctorate on Artificial Intelligence run by Sapienza University of Rome in collaboration with the University of Cagliari.
This work was partially supported by project SERICS (PE00000014) under the NRRP MUR program funded by the EU - NGEU.

\bibliographystyle{unsrt}
\bibliography{reference}

\end{document}